\documentclass[preprint]{aastex}
\newcommand{\kms}{km~s$^{-1}$}
\newcommand{\vsini}{$v\sin i$}
\newcommand{\msun}{M$_{\sun}$}
\newcommand{\sigmag}{$\beta_{\rm Gray}$}
\shorttitle{Rapid Rotation of K Giants}
\shortauthors{Carlberg et al.}

\begin{document}

\title{The Frequency of Rapid Rotation Among K Giant Stars}

\author{Joleen K. Carlberg\altaffilmark{1}, Steven R. Majewski\altaffilmark{1}, Richard J. Patterson\altaffilmark{1}, Dmitry Bizyaev\altaffilmark{2,}\altaffilmark{3}, Verne V. Smith\altaffilmark{4},
Katia Cunha\altaffilmark{4} }
\altaffiltext{1}{Dept. of Astronomy,
University of Virginia,
Charlottesville, VA 22904
 jkm9n@virginia.edu, srm4n@virginia.edu, ricky@virginia.edu}
\altaffiltext{2}{Apache Point Observatory
Sunspot, NM
 dmbiz@apo.nmsu.edu}
\altaffiltext{3}{Sternberg Astronomical
Institute, Moscow, 119992, Russia}
\altaffiltext{4}{NOAO
Tucson, AZ
 vsmith@noao.edu, cunha@noao.edu}

%250 words max on abstract!
\begin{abstract}
We present the results of a search for unusually rapidly rotating giant stars in a large sample of K giants ($\sim$1300 stars)
 that had been spectroscopically monitored as potential targets for the Space Interferometry Mission's Astrometric Grid.
The stars in this catalog are much fainter  and typically more metal-poor than those of other catalogs of red giant star rotational velocities, but the spectra generally only have signal-to-noise (S/N) of $\sim$20--60, making the measurement of the widths  of individual lines difficult.  To compensate for this, we have developed a  cross-correlation method to 
 derive rotational velocities in moderate S/N echelle spectra to efficiently probe this sample for rapid rotator candidates. 
We have discovered 28 new red giant rapid rotators as well as one extreme rapid rotator with a \vsini\ of 86.4~\kms.   Rapid rotators comprise 2.2\% of our sample, which is  consistent with other surveys of brighter, more metal-rich K giant stars.
Although we find that the temperature distribution of rapid rotators is similar to that of the slow rotators,  this may not be the case with the distributions of surface gravity and metallicity. The rapid rotators show a slight overabundance of low gravity stars and as a group are significantly more metal-poor than the slow rotators, which may indicate that the rotators are tidally-locked binaries. 
\end{abstract}

\keywords{stars: late-type, stars: rotation}
\section{Introduction}
\label{sec:intro}
Rotation is an important stellar physical property that plays a role in a star's magnetic activity, circulation of convective material,  and overall internal structure.  Rotation may even give clues to a star's dynamical history---e.g., evidence for an interaction with a stellar or sub-stellar companion. 
 Early surveys of projected rotational velocity,  \vsini,  in giant stars led to the discovery that giants cooler than about 5000~K are predominately slow rotators \citep{gray81,gray82}, and later work by \cite{deMed96b} confirmed that cool giants are characterized by $v \sin i \leq 2$~\kms.    Reduced rotation speeds are expected in giant stars because, due to angular momentum conservation,  they should spin down as their radii expand.
 Efficient magnetic braking \citep{gray81,gray82} also ensures the ubiquity of slow rotation speeds.

 The few percent of K giant stars that are  rapidly rotating   are therefore interesting because their envelopes must have been 
 augmented with angular momentum.   Some of these rapid rotators may have unseen binary companions with which they have interacted. Other rapid rotators are seemingly isolated, and   \cite{simon89} proposed that these  isolated rapid rotator giants are going through a short-lived rapid rotation phase caused when the deepening stellar convection layer dredges up angular 
momentum from the more rapidly rotating stellar interior.
Alternatively, a sub-stellar companion, such as a brown dwarf or planet, could also impart excess angular momentum to the stellar envelope 
as tidal forces with the primary decay the substellar companion's orbit.
\cite{pete83} were the first to consider planets as sources of angular momentum in evolved stars;  subsequent discoveries of  Jupiter-mass planets 
orbiting their stars at inner-Solar System distances  lent further 
credibility to this idea because these planets are close enough to be eventually accreted by their host stars yet  massive enough to impart significant angular momentum 
\citep[see, e.g., ][]{siess99,livio02,soker04,massarotti08a,carlberg09}.

The study of rapid rotators is clearly limited to those stars for which \vsini\ have been measured, and most large surveys of \vsini  \ in giant stars still focus on relatively bright giants.  For example, \cite{demed99} chose targets  from ÒThe Bright Star CatalogÓ \citep{hoffleit82}; their survey is complete to $V=6.3$ mag, while the more recent \vsini\ catalog of Hipparcos giants
by   \cite{massarotti08a}  surveyed stars with, on average,  $V=6$, and the faintest star at $V=9.5$.
In contrast, we aspire to measure \vsini\ on a 
relatively faint ($\overline{V}= 9.7 $) sample of stars  for which moderate signal-to-noise (S/N~=~20--60) echelle spectra already exist
as part of a radial velocity monitoring program on behalf of the Space Interferometry Mission (SIM).  The average increase in survey sensitivity of 3.4 magnitudes extends the distance over which 
we can find rapid rotators by a factor  of 4.8,  corresponding to increasing the volume surveyed by over two orders of magnitude.  
 Such a catalog would extend our understanding of rapid rotators to stars that are likely more metal 
poor and older, as well as in  different Galactic environments, than those rapid rotators that have been studied thus far.

 Ideally, spectra with both high spectral resolution, which determines the minimum measurable rotational velocity, and high S/N are desired to measure \vsini.  Both of these requirements increase the necessary observing time for a given apparent magnitude.  Because integration times at fixed S/N increase roughly as the square of the decreasing flux from an object, the observational requirements become prohibitively time-consuming for faint objects.  To circumvent the stringent S/N requirements, we developed an   
 approach whereby very accurate measures of \vsini \ from high S/N spectra of the brightest stars lay the groundwork for empirically calibrating indirect measures of \vsini\  made with lower S/N spectra.
  Although the accuracy  of the measured \vsini \ will be lower in the latter spectra, the trade-off is the ability to derive \vsini\ for many fainter stars more quickly than with
  more accurate but observing-expensive \vsini \ measurements.

We describe our sample of K giants in \S \ref{sec:sample} below.  Our method for measuring projected rotational velocities is described in \S \ref{sec:method}, and our
results on the frequency of rapid rotators in K giants are presented in \S \ref{sec:results}.   Our discussion in   \S \ref{sec:binary}  addresses the effects of potential undetected binaries in our sample, while \S \ref{sec:discuss} covers the implications  of our results for the proposed spin-up mechanisms that could create   K giant rapid rotators.

\section{Sample: Astrometric Grid K Giants}
\label{sec:sample} 
The giant  stars comprising our sample were  observed for radial velocity stability and characterization by \cite{dmbiz06,bizyaev10}
as candidates for SIM's Astrometric Grid \citep{ciardi04,catanzarite04}; we use these same spectra here for the \vsini\ measurements.
The grid candidates were originally selected from
both the Tycho catalog \citep{hoeg00} and from the Grid Giant Star Survey \citep[GGSS,][]{ricky01}, the latter of which was designed
explicitly to find suitable astrometric grid candidates. These two sources differ slightly in their coverage of stellar metallicity, brightness,  and
temperature. The GGSS stars are on average more metal-poor and fainter,  and they cover a broader range of
temperatures than the Tycho stars.   
In Table \ref{tab:alias}, we list our program stars and their known aliases in other major catalogs including  Hipparcos, Henry Draper, Bonner Durchmusterung,  and Tycho. 
When referring to our program stars, we prefer to keep the GGSS designations even when Tycho aliases exist because the names indicate how the stars were originally selected. 

The total sample presented in this work  is composed of 613 GGSS and 684 Tycho stars.  
 All 1297 of our stars have photometric temperatures (Washington $M-T_2$ for GGSS and $B_{\rm t}-V_{\rm t}$ for Tycho), all of the GGSS stars have photometric metallicities from Washington/DDO51 colors \citep[e.g.,][]{majewski00}, and 997 stars have spectroscopically derived temperatures, metallicities, and surface gravities from the pipelines described in \cite{dmbiz06,bizyaev10}, hereafter B06 and B10.  The pipeline in B10 also produced  \vsini\ estimates; however, their pipeline works best on the slow to moderate rotators and has difficulties constraining \vsini\ for the more rapid rotators.  We provide an independent \vsini\ for these 842 stars as well as the first reported \vsini's for the 455 stars {\it not} overlapping B10.   For the 155 of our stars  overlapping with the B06 sample we also have radial velocity dispersions. Our sample is characterized by stellar temperatures ranging from 3750-5850~K with an average of 4540~K, and has metallicities ranging 
from $-2.8$~dex to $+0.6$~dex with an average of $-0.3$~dex. % [[Reddening]

The observations were carried out with the 2.1-m telescope at McDonald Observatory with the Sandiford Cassegrain Echelle spectrograph between January 2001 and July 2006.
The spectra have  high nominal resolution ($R \sim$ 55,000)  and a typical S/N between 20 and 60.  
Standard IRAF tasks were used to reduce the data; wavelength calibrations were made with Th-Ar lamp spectra  that immediately preceded each program star observation.
 (A more complete discussion of this sample can be found in B06  for the ``Northern Sample'' of that paper and in B10.)
These stars are either first ascent red giant branch (RGB) stars,  asymptotic giant branch (AGB) stars, or red clump stars; all three types of stars can fall in  the same temperature
 and surface gravity space probed by our sample.  

\section{Projected Rotational Velocities}%{Method}
\label{sec:method}
The projected rotational velocities of the stars are measured by analyzing the 
stellar line widths via a cross-correlation technique using a high resolution template.  
The steps leading from the cross-correlation peak width to \vsini\ are described fully in the paragraphs below and are summarized as follows:
(1)  measure the cross-correlation peak width ($\mu$) from each echelle order,  (2) subtract from $\mu$ the contribution from the template  ($w_{\rm t}$) to get the 
  line width of the target star ($w_{*}$) for each echelle order, 
(3) convert $w_{*}$ to total stellar broadening ($\beta_{\rm Gray}$) and average all values from different echelle orders and, when applicable, different images, 
 and (4) separate $\beta_{\rm Gray}$ into physical broadening components (macroturbulence, rotation, etc.).

The high-resolution Arcturus spectrum of \cite{hinkle00}  was chosen 
as the cross-correlation template.  For each star in our sample, we create an echelle template by subdividing the Arcturus spectrum  into ``pseudo-orders'' and 
re-sampling to match the wavelength coverage and pixel resolution of the echelle orders of the target star spectra.  
Each target star spectrum is then cross-correlated with its matching template. \cite{TRD79} 
found that the width of the resulting cross-correlation peak, $\mu$, could be described as the quadrature sum of the typical width of the spectral lines in the template
spectrum, $w_{\rm t}$, and the typical width of the spectral lines in the object spectrum, $w_{\rm obj}$, i.e.,  $\mu^{2}=w_{\rm obj}^{2}+w_{\rm t}^{2}$.
We measure $\mu$ by fitting a Gaussian function to the cross-correlation peak, and each template order is cross-correlated against itself to measure $w_{\rm t}$. 
(In this special case,  $\mu^2=2w_{\rm t}^2$.)

The typical width of the object spectrum can itself be modeled by the quadrature sum of the total stellar width and
the instrumental width, $w_{\rm obj}^2=w_*^2+w_{\rm i}^2$.  
The instrumental width for each echelle order is measured from the ThAr calibration lamp images. 
We find that the instrumental width is frequently larger than the 5~\kms\ expected for the McDonald 
echelle\footnote{http://www.as.utexas.edu/mcdonald/facilities/2.1m/ce.html}, with $w_{\rm i}$ ranging generally from 5-11~\kms\ and having both an average 
and median of 7.0~\kms.  A likely cause of this discrepancy is a focusing variation across observing runs.  

After subtracting the instrumental width, we are left with the average width of the spectral lines, $w_*$, due only to physical stellar processes. 
\cite{fekel97} calibrated an empirical relation between the easily measured average FWHM of spectral lines (in \AA\ after removing the instrumental
broadening) and the total stellar broadening %NEW TEXT
 ($\beta_{\rm Gray}$, measured in ~\kms).  This $\beta_{\rm Gray}$ is the quadrature sum of the \vsini\ and macroturbulence ($\zeta$) values measured by \cite{gray82, gray89} using Fourier analysis.
This relationship was  established using a set of a half  dozen lines centered roughly at $\lambda=6455$~\AA, 
which is given by their equation 1.  To make their equation applicable to FWHM measurements made at any 
wavelength and in velocity units, we multiplied their polynomial coefficients by $c/\lambda$ to obtain    
\begin{equation}
%ORIGINAL: 0.0408200    0.0250900  0.000140000
%MODIFIED: 1.8958186    1.1652643  0.0065020730
w_*({\rm km}\  {\rm s}^{-1}) = 1.89582 + 1.16526\cdot\beta_{\rm Gray} + 0.00650\cdot\beta^2_{\rm Gray}
\label{gray}
\end{equation}
Using equation (\ref{gray}), we calculate $\beta_{\rm Gray}$ for each echelle order, resulting in $\sim$24 independent measurements 
of the stellar broadening for each stellar spectrum. Values of $w_* <$ 1.896~\kms\ would result in a negative value for the total stellar broadening; 
for these echelle orders we set the stellar broadening to zero.  

\begin{figure} %F1
\includegraphics[scale=0.7]{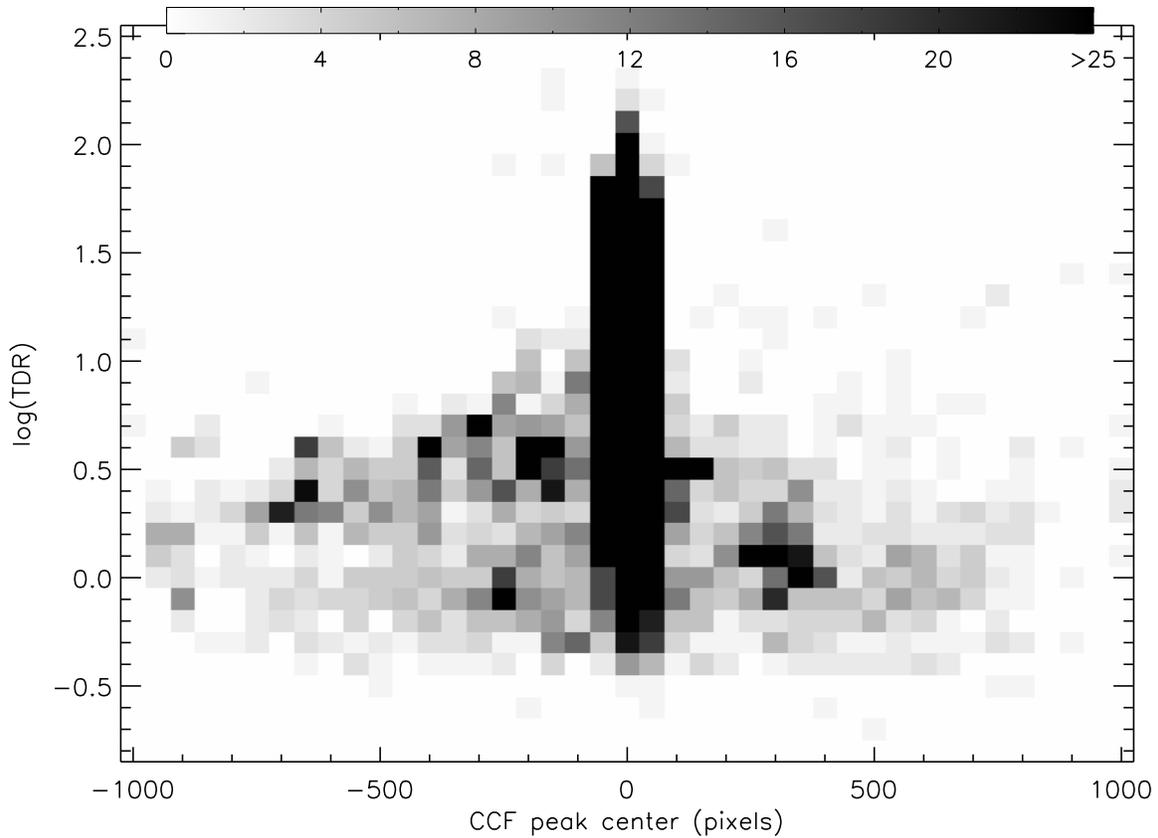}
\caption{Number density of stellar orders having a given TDR and center of the cross-correlation function peak.  The color bar ranges from 0 to $25$. 
Note that most of the orders with $|{\rm center}| > 100$, which are likely noise peaks,  can be removed by requiring $\log({\rm TDR}) > 1$.
\label{fig:tdr}}
\end{figure}

At this point, we can average {\it all} measurements of $\beta_{\rm Gray}$ for each star, including
all those taken for the same star on different nights.  However, some echelle orders cross-correlate better with the template than others, and we use the 
Tonry-Davis Ratio\footnote{The TDR measures the ratio of the true cross-correlation peak to the average height
of the noise peaks.} \citep[TDR;][]{TRD79}, which is computed by IRAF's {\it fxcor} task, to cull these measurements.
We require TDR~$> 10$ for an individual \sigmag\  measurement to be kept.
The value of this cut-off was guided by the distribution of  TDR with the pixel offset of the cross-correlation peak---zero pixel offset indicates zero shift between the centers of the spectral 
lines in the target star compared to the template.    Non-zero pixel offsets are expected because the target stars have a radial velocity; however, excessively large pixel offsets (either positive or negative) suggest that 
those cross-correlation peaks are in fact noise peaks.   In  Figure \ref{fig:tdr}, we show the distribution of pixel offset and TDR  for all of our spectra.
By selecting orders with TDR greater than 10, we remove all orders where the selected peaks are obviously noise based on their pixel offsets. 
For stars with no echelle orders meeting this TDR requirement,  all the measurements were used, but the final result is flagged as being potentially unreliable. 
The final measurement of the stellar broadening
for each star is computed by an iterative sigma-clipped average of the individual measurements to discard outliers (3$\sigma$ away from the median
value in each iteration). 
The number of measurements of  $\beta_{\rm Gray}$ kept for each star (all orders {\it and} spectra) ranges from 3 to 86 with a mode of 22, and we 
use the standard error in the mean to gauge the uncertainty in the stellar broadening.  

The final step is to disentangle the major physical broadening mechanisms that contribute to \sigmag. The main broadening contributors in giant stars
 are the macroturbulence  and, when large, the rotational velocity. The macroturbulence increases with stellar 
temperature, and we use photometric colors to derive the temperature 
using the Cousins $V-I$ and Johnson $B-V$ color-temperature relations of \cite{houd00} for the GGSS and Tycho stars, respectively.
The GGSS stars have Washington $M-T_2$ colors, which are converted to Cousins $V-I$ \citep{majewski00}. Similarly, the Tycho  stars have $B_{\rm t}-V_{\rm t}$, which are converted to Johnson $B-V$.
 These photometric temperatures are then used to estimate
macroturbulence using the formula given in \cite{Hekker07} for class III giants. The projected rotational velocity, \vsini,  
is finally found by subtracting out the macroturbulence contribution to the stellar broadening, i.e., 
\begin{equation}
v\sin i=\sqrt{\beta^2_{\rm Gray}-\zeta^2},
\label{width3}
\end{equation}
  as was done in both \cite{fekel97} and \cite{Hekker07}.
For thirty stars, the estimated macroturbulent broadening was larger than the measured stellar broadening; in the plots and tables that follow we display these
stars as having  an upper limit of  \vsini~$<$~3~\kms.

 All of the stars  that were not flagged in the \vsini\ derivation are listed in Table \ref{tab1} along with their respective right ascension, declination, \vsini, error in \vsini, number of individual measurements
contributing to  \vsini, and photometric temperature.  
Twenty stars were flagged during the \vsini\ derivation by failing to have any echelle orders meet the TDR requirement.
For seven of these flagged stars (G0224+05.38, G0439+05.244, G1038-11.106, G1817-16.1565, G1856+05.2998, 
G1856+05.8646, and G1955+28.13080), the spectra simply have too low a S/N to get a good cross-correlation peak, and these stars are removed from the sample. 
The remaining 13 stars are presented in Table \ref{tabFL}, which lists the same set of stellar properties as Table \ref{tab1}.
We address flagged stars separately in the following plots and discussion to emphasize the larger degree of uncertainty in their measured stellar broadening.

\section{Results}
\label{sec:results}
\begin{figure} %F2
\includegraphics[scale=0.7]{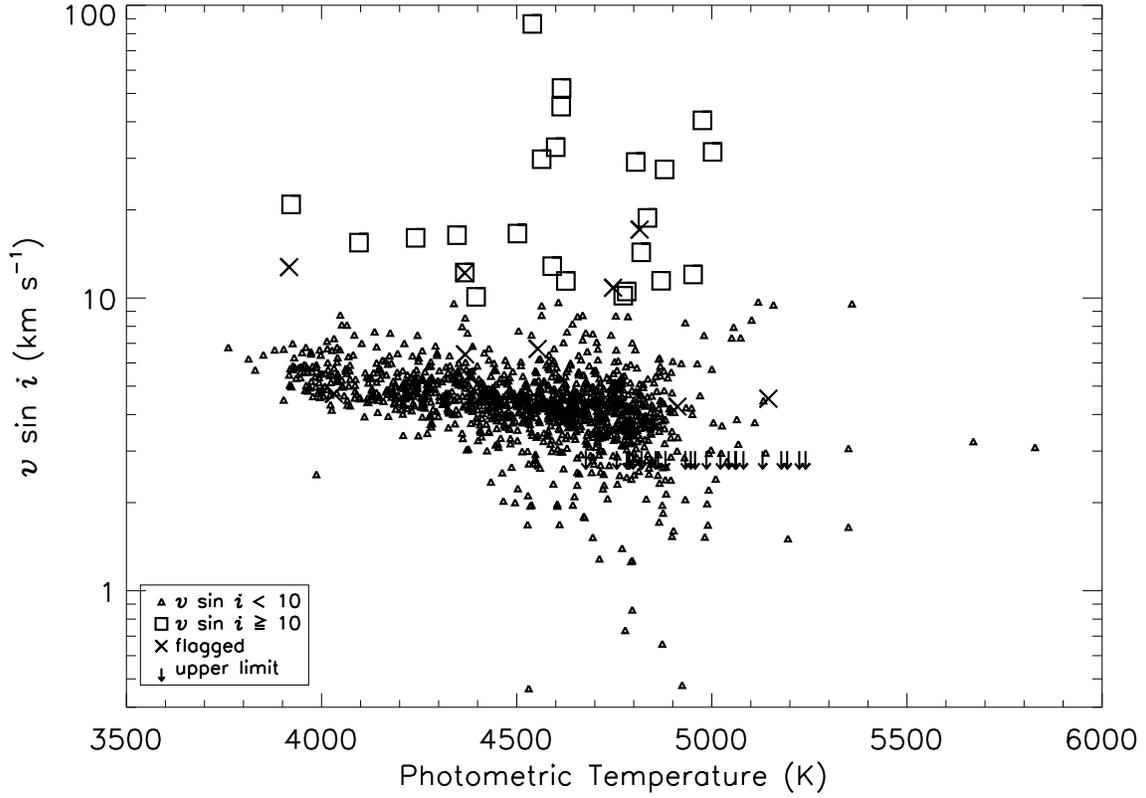}
\caption{Projected rotational velocity, \vsini, as a function of temperature for all of the stars in our sample. 
Triangles, squares, and X's represent slow rotators (\vsini\ $<$ 10~\kms), rapid rotators (\vsini\ $\ge$ 10~\kms), and
flagged stars (see Section \ref{sec:method}), respectively.  Arrows indicate upper limits. The typical error bar in \vsini\ is 0.5~\kms.
\label{rrs}}
\end{figure}

\begin{figure}  %F3
\includegraphics[scale=0.7]{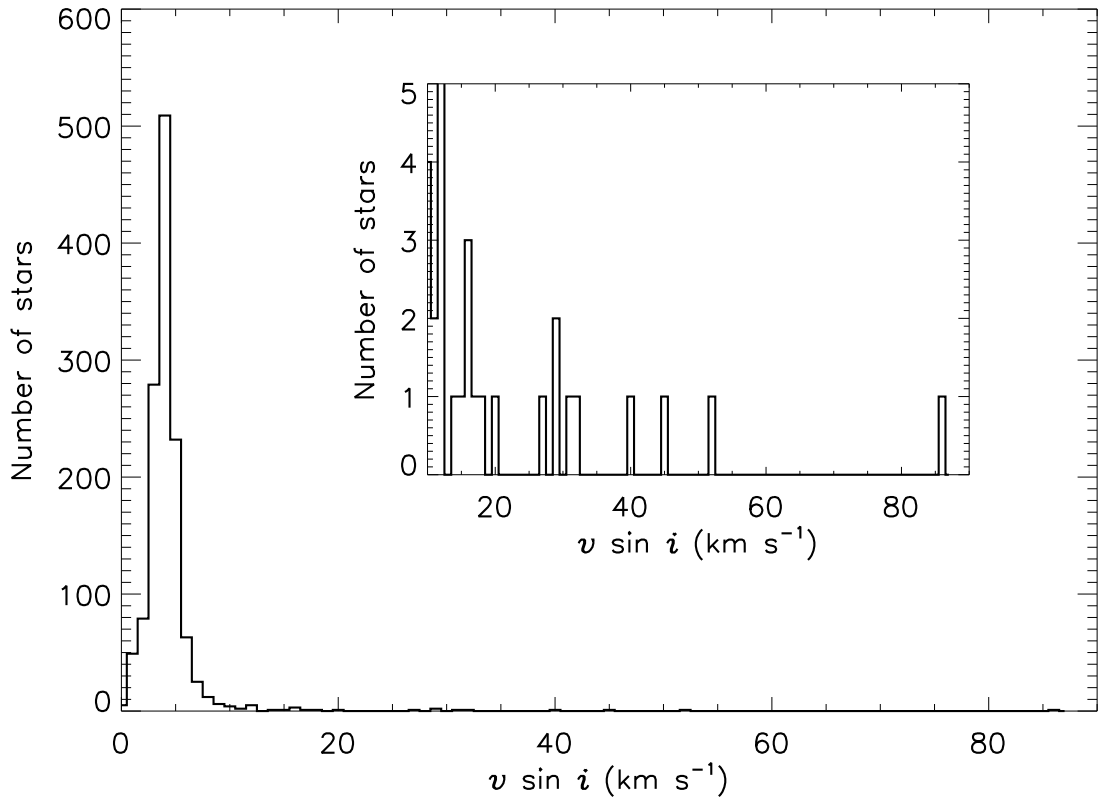}
\caption{Histogram showing the  \vsini\  distribution of our giant stars.  The inset shows just the rapid rotation region (\vsini\ $\ge$ 10~\kms) for clarity.
\label{rrs_hist}}
\end{figure}

In Figure \ref{rrs} we plot the measured rotational velocities as a function of photometric temperature.  
The slight decreasing trend of \vsini\ with increasing temperature, as well as the increasing scatter in \vsini,  are due to the way that uncertainties in $\zeta$ propagate to \vsini.  
Because $\zeta$ is subtracted in quadrature, a constant uncertainty in $\zeta$ will yield larger \vsini\ errors at larger $\zeta$ (and thus at higher temperatures).
This behavior accounts for the increased \vsini\ scatter at higher temperature.
Additionally, at constant $\zeta$, a  positive  error in $\zeta$ results in a larger \vsini\ error than a negative error of the same amount; this accounts for the slight decreasing trend of \vsini\ with temperature.
We define rapid rotators to be those giant stars with a rotational velocity of 10~\kms\ or larger.  This cut-off is the same one that  is used by \cite{fekel97},  and it  is slightly more conservative than the cut-off made by \cite{drake02}, who used 8~\kms, and \cite{massarotti08a}, who identified three non-binary rapid rotators with \vsini$=7-10$~\kms.  %7.7, 8.4, and 9.9
 A total of 28 of our stars qualify as rapid rotators by our definition, 24 of which are not flagged in 
the \vsini\ analysis. This translates to 2.2\% of the giants being rapid rotators if all stars are included in the sample, or 1.9\%
if all flagged stars are ignored.   In Figure \ref{rrs_hist}, we show a histogram of the \vsini\ distribution; the rapid rotation portion is shown in the inset for clarity.
In addition, we have discovered one extreme rapid rotator, Tyc5854-01526-1, which has \vsini~$\sim 86.4$~\kms.  Of the catalogs mentioned in Section \ref{sec:intro}, 
Tyc5854-01526-1's \vsini\ exceeds all those in \cite{demed99} and all but three stars in \cite{massarotti08a}.

\begin{figure} %F4
\includegraphics[scale=0.7]{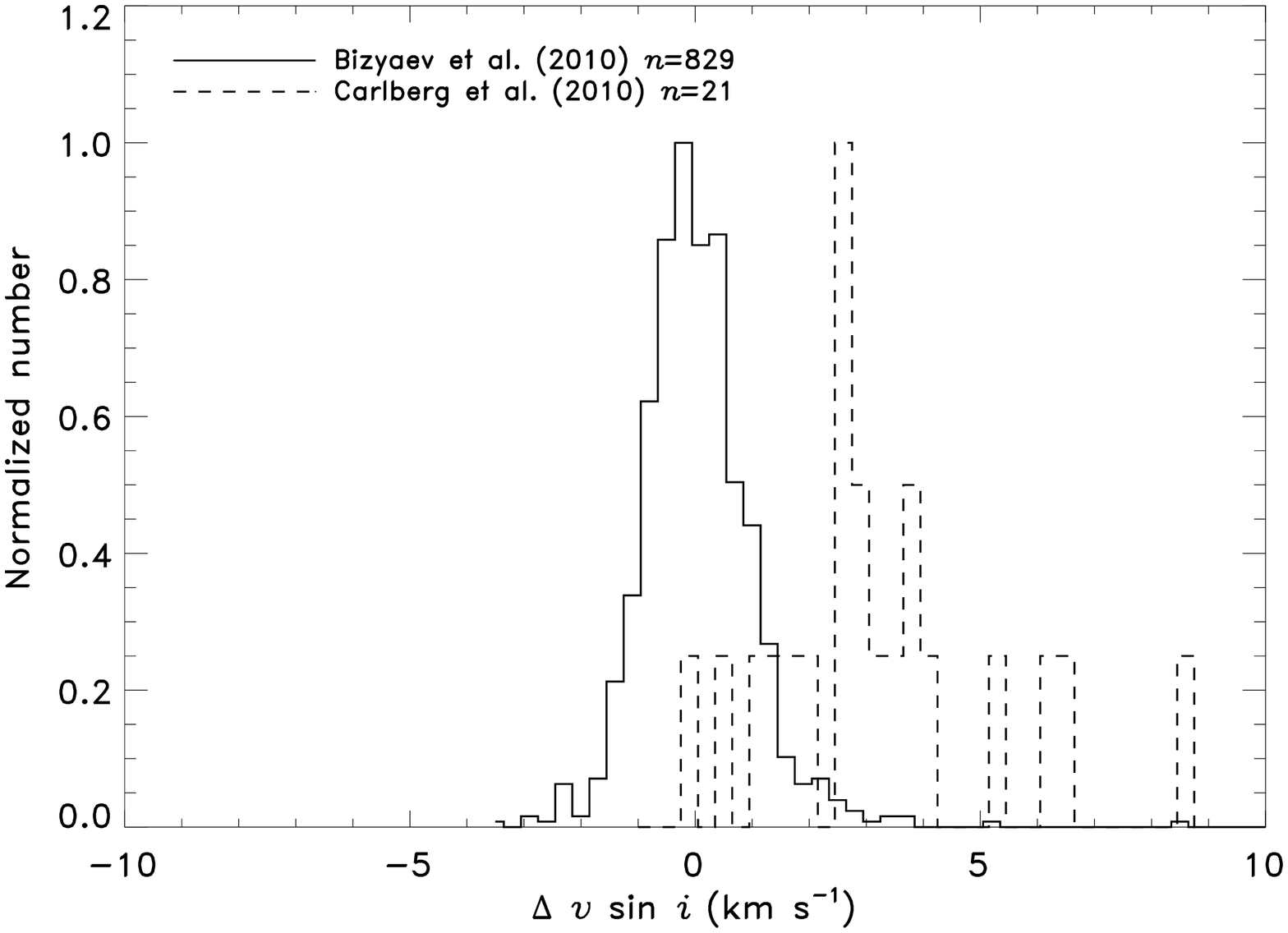}
\caption{\label{vsini_comp_both} Comparison of rotational velocities between this work and those derived by \cite{bizyaev10} from the same data (solid line) and \cite{carlberg10} from high S/N data (dashed line), shown as normalized histograms. The $\Delta$\vsini\ is defined as \vsini\ [this work]$-$\vsini~[literature]. Most of the B10 measurements deviate from our measurement by less than 2~\kms, although  there are four stars that are outside the range of this plot.  The \vsini\ measurements from \cite{carlberg10} tend to be smaller than our measurements by 2.7~\kms\ on average.}
\end{figure}

The \vsini\ values can be compared to those calculated with the pipeline described in B10. 
In that pipeline,
\vsini\ was the last stellar parameter to be measured, and it was found by fitting synthetic spectra of varying \vsini\ to a single echelle order centered roughly at
 $\lambda5740$~\AA.  
In Figure \ref{vsini_comp_both}, we plot a normalized histogram of the differences between these two derivations of \vsini.   For this subset of stars, which is 64\% of our sample, both versions of the rotational velocity were derived from the same spectra.  For clarity, the scale of the x-axis excludes four stars with large differences between the two \vsini\ measurements.    Over 94\% percent of the stars have our \vsini\ measurements agree with the alternative measure of \vsini\ by B10 to within  2~\kms.   
The \vsini\ values derived here can also be compared with \vsini\ measurements obtained from high S/N follow-up spectra
observed as part of a project to look at abundance differences between slow and rapid rotators \citep[see, e.g.,][]{carlberg10}.  These spectra have $R \sim 31,500$ and $R \sim 43,000$ for observations taken with the Apache Point Observatory 3.5-m and Kitt Peak 4-m telescopes, respectively, and all have S/N~$\sim 100$ per pixel.
The \vsini 's were measured by fitting synthetic spectra to six isolated iron lines to fit the broadening.  
The difference between these high S/N \vsini\ measurements and our measurements are also plotted as a normalized histogram in Figure~\ref{vsini_comp_both}, and we find that our \vsini\ measurements tend to be larger  by about  2.7~\kms.   Because the cross-correlation peak measures the average width of all the lines in the spectral order,  it is likely that this slight overestimate of \vsini\ comes from blended absorption lines in the spectra.  Alternatively, an overestimate of either the instrumental or macroturbulent broadening in the \cite{carlberg10} data could result in a systematic underestimate of their \vsini's. Nevertheless, we conclude from Figure \ref{vsini_comp_both} that our cross-correlation approach works well enough to identify rapid rotators.  

\begin{figure} %F5
\includegraphics[scale=0.8,angle=90]{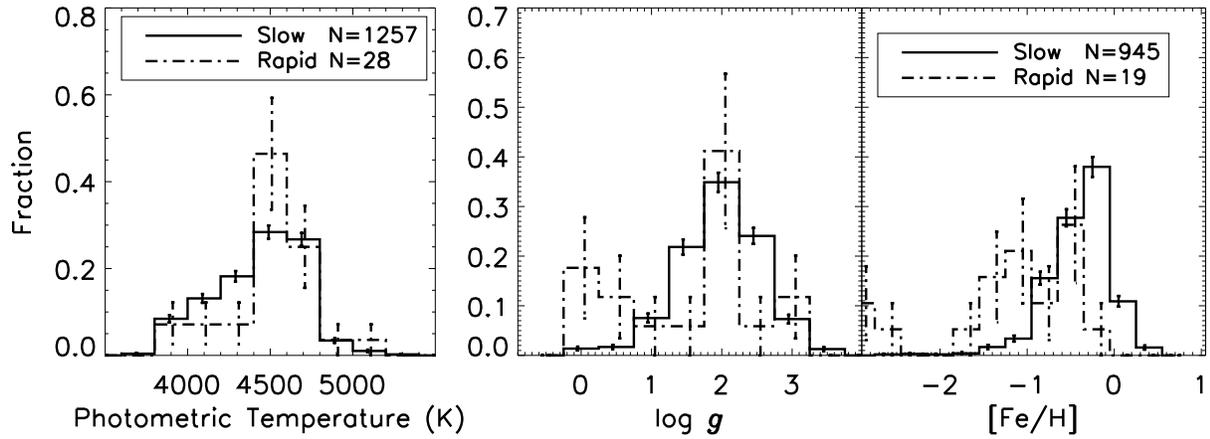}
\caption{The distribution of normal stars (solid lines) and rapid rotators (dot-dash lines) for: 
({\it left})  photometrically derived effective temperatures in bins of 200 K, ({\it middle})  $\log g$ in bins of 0.5 dex, and 
({\it right}) metallicity in bins of 0.3 dex.  Error bars are Poisson $\sqrt{N}$ errors and are shown offset from bin center for clarity. \label{histt1}
These histograms {\it include} the flagged stars.}
\end{figure}

Our large sample of K giants allows a statistical study of the coincidence of rapid rotation with other basic parameters of the star. 
Recall from Section \ref{sec:sample} that a subsample of 997 stars have the spectroscopically-derived stellar parameters from B06 and B10; we use this subsample to test for correlations. Figure \ref{histt1} shows the distribution of normal rotators and rapid rotators in bins of photometrically-derived temperature, and spectroscopically-derived $\log g$ and metallicity.  
The distribution of rapid rotators in temperature looks similar to that of the normal rotators; the two samples peak in the same bin, at  4500 K,  and have roughly the same shape.

In   $\log g$, the rapid rotators show an overabundance of stars in the smallest $\log g$ bin, and these stars are likely the most luminous stars. 
 It is possible that the macroturbulence in these stars has been underestimated (leading to an overestimated \vsini) because at a given stellar temperature $\zeta$ is larger for brighter luminosity classes \citep{Hekker07}.
  However, since the three lowest surface gravity stars all have \vsini\ exceeding 15~\kms, their measured rotation will still be rapid even if a larger $\zeta$ contribution is removed. In a study of  metal-poor giants,  \cite{carney03} similarly found that a relatively large fraction (20--30\%) of their most luminous single giant stars are rapidly rotating. Their finding is consistent with our findings since the  
three rapid rotators in our sample with smallest $\log g$ are metal-poor ([Fe/H]$< -1.0$).   \cite{massarotti08a} also finds relatively enhanced rotation speeds in the most luminous red giants in their sample as well as enhanced rotation in horizontal branch stars.
 
In the metallicity distribution,  there is a pronounced shift in the median abundance towards lower metallicities for the rapid rotators:
while the samples peak in adjacent bins---centered at $-0.5$ dex for the rapid rotators and $-0.2$ dex for the slow rotators---there is a pronounced under-abundance of rapid rotators in the more metal-rich bins as well as a second peak in the distribution at $-1.2$ dex.
In particular, only one rapid rotator, at [Fe/H]~$=-0.18$,  falls in the bins at or more metal-rich than the peak of the slow rotators.
The mean [Fe/H] is $-1.1$ dex for the rapid rotators and $-0.3$ for the slow rotators.
A KS-test of the metallicity distributions of 
the normal and rapid rotators shows that there is only a 3.5$\times 10^{-4}$\% chance of these samples being drawn 
from the same parent sample.   
Because rapid rotation  dramatically broadens and reduces the maximum depth of the absorption lines, it is
possible that the stellar parameter pipeline of B10  underestimates the metallicity of the 
rapid rotators.  To check this hypothesis, we artificially broadened spectra of normal rotators and rederived  the
stellar parameters.  We did find that the addition of fast rotation will decrease the derived metallicity by about 0.2 dex; however,
this is far from being able to explain a 0.7 dex offset seen between the averages of these two samples.   ``Correcting'' the rapid rotator
metallicities by 0.2 dex brings the KS-probability up to  only 0.03\% and cannot close the gap between the mean metallicities of the two samples.
Thus, there seems to be an intrinsic metallicity distinction between the slow and rapid rotator samples.

While a thorough discussion of the implications of this metallicity distinction is beyond the scope of this paper, we briefly mention a few.    First, both metallicity and rotation alter the internal structure of stars, and each property can influence the other.  For example, \cite{maeder01} found that in massive stars (more massive than the stars in this study), a low metallicity can inhibit  mass loss and, by extension, angular momentum loss.  For stars with the same initial rotational velocity, such a relationship would lead  to larger rotations in the metal poor stars later in their evolution.
Alternatively, the relationship between metallicity and rotation may also be the result of the rapid rotation formation mechanism; possible mechanisms are discussed in detail in Section \ref{sec:discuss}.

\section{Binary Contamination}
\label{sec:binary}
Because we are interested in studying the rotational speeds of single giant stars, we are naturally concerned with binary stars potentially contaminating our sample.  Therefore, we attempted to match our program stars with published catalogs of binaries with red giant primary stars.  A cross-reference of
our target list with \cite{demed99}, \cite{demed00}, \cite{demed02}, and \cite{massarotti08a} yielded no matches.
 A coordinate  search of the ``9th Catalog of Spectroscopic Binary Orbits'' \citep{pourbaix04}
yielded only one match for Tyc3340-01195-1.    The binary orbit solution  for this star originally comes from \cite{mermilliod07}---identified as NGC~1528~4 in that study---and lists a period of 1071 days.

The lack of cross-referencing matches means that the majority of the stars have not had their radial velocities monitored long enough to detect binaries, necessitating  a statistical analysis  of possible binary contamination.
Binary stars can contaminate our rapid rotator sample if the spectral features of the two stars add in such a way that there appears to be only one set of broadened stellar
absorption lines as opposed to two sets of narrow absorption lines.  In this case, not only is the star not an isolated star, but it also not a rapid rotator. 
For this contamination  to happen, the companion star must contribute a significant enough amount of flux, at  least  comparable to the noise.   Our best spectra 
have S/N~$\sim 60$, which means companion stars that have as little as
2\% the flux of its primary will be ``visible'' over the noise.  This corresponds to a difference of  about 4 magnitudes.  
Using \cite{giard00} isochrones, we estimate the approximate companion mass, $M_2$, of the faintest stars still capable of contributing significantly to the flux, expressed
in terms of the mass ratio $q$, where $q=M_1/M_2$.   For RGB stars between ages 0.1 and 14 Gyr and metallicities between $-1.7$ and +0.17 dex, 
we find generally that $q_{min}=0.7$.
We then use the distribution function of $q$ presented in  \cite{duquennoy91} to estimate the fraction of binary stars with companions between
$q=0.7$ and $q=1$  to be  $9.5\%$.

\begin{figure} %F6
\includegraphics[scale=0.7]{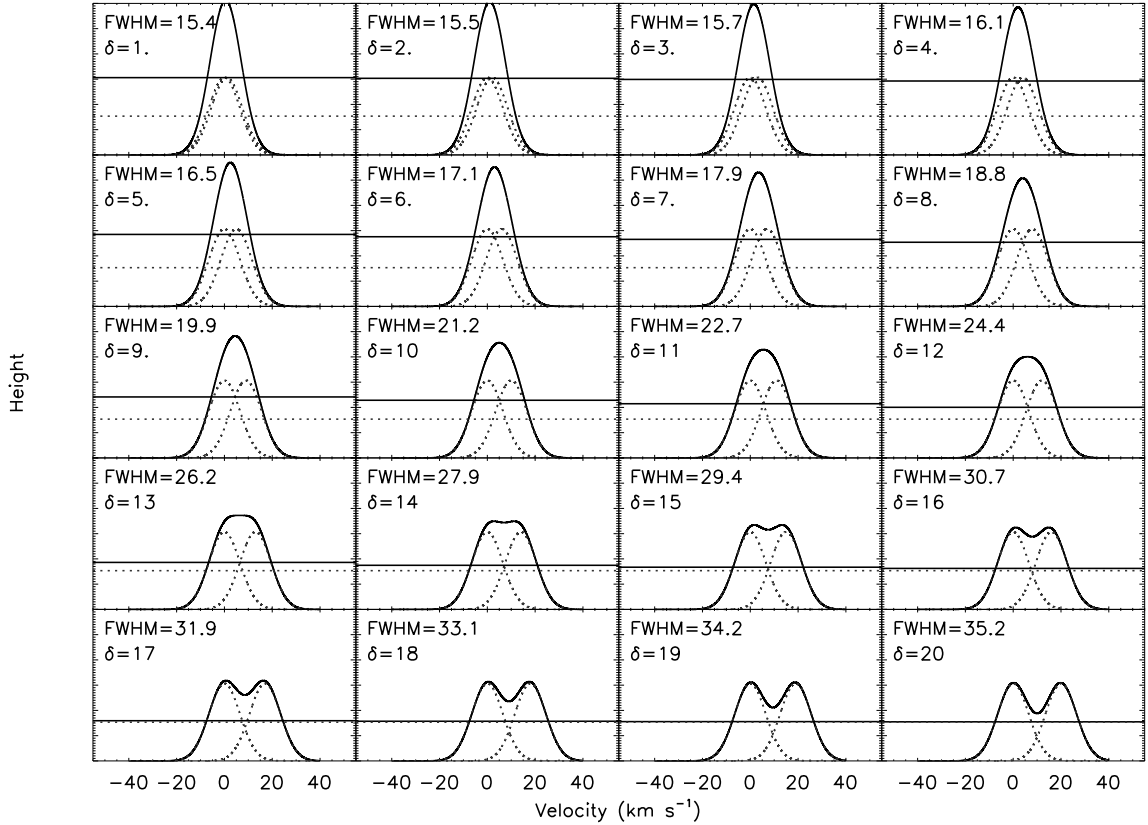}
\caption{Modeled cross-correlation peaks of a binary spectrum with stars of equal brightness, represented by the sum of two Gaussian components of equal width and height.
 The component Gaussians are shown in each panel with dotted lines, and they each have  $\sigma=6.5$~\kms\ (FWHM $\sim 15.3$~\kms).  The horizontal lines show the 
 half-maximum level of the sum (top solid line) and the components themselves (bottom dotted line).  Listed in each panel is  the FWHM of the sum and the
velocity separation between the components, given by $\delta$; both have units of~\kms.  Peaks with FWHM of 21.9~\kms\ or higher, which occurs beyond $\delta$ of $\sim$10.5~\kms,  would have  measured \vsini~$> 10$~\kms, and would thus appear to be rapid rotators.  Note that distinct peaks are seen for $\delta \ge 15$~\kms, and these
stars should clearly indicate the presence of a double-lined spectroscopic binary.
\label{blends}}
\end{figure}

Not only must the companion star be bright enough to contribute significantly to the light, but the lines of the two components must be offset in velocity such that at our
spectral resolution the sets of absorption lines of the two stars blend to mimic a single set of broad lines.  This broad-line signature would also appear in the cross-correlation peaks.
We created a simple model of this situation by assuming that the cross-correlation peak from a spectrum of such a binary system would look like the sum of two Gaussian components, one for each
star of the binary.  The largest effect would occur for  two stars of equal brightness and temperature. 
In Figure \ref{blends}, we show the sum of two equal height Gaussian components, each with a FWHM of 15.3~\kms, which is approximately the median  value
of the cross-correlation widths ($\mu$,  see Section \ref{sec:method}) that we measured in the normally rotating K giants in our sample.  The model binary components are separated in velocities by an offset, $\delta$, ranging between 1-20~\kms.
Working backwards through our \vsini\ derivation described in Section \ref{sec:method} (assuming $T=4500$~K to derive $\zeta$ and using median values of $w_{\rm i}=7$~\kms\ and $w_{\rm t}=13.68$~\kms) we find that a cross-correlation peak with FWHM larger  than 21.9~\kms\ would have a derived \vsini \ greater than 10~\kms.   This condition is met in the blended profiles of Figure \ref{blends} for $\delta \gtrsim 10.5$~\kms. These binaries would appear to be single rapid rotators, whereas, for $\delta \ge$ 15~\kms, the components begin to show distinct peaks.  
For $\delta < 10.5$~\kms, the stars would appear to have enhanced rotation, but not enough to be classified as a rapid rotator.

\begin{figure} %F7
\includegraphics[scale=0.7]{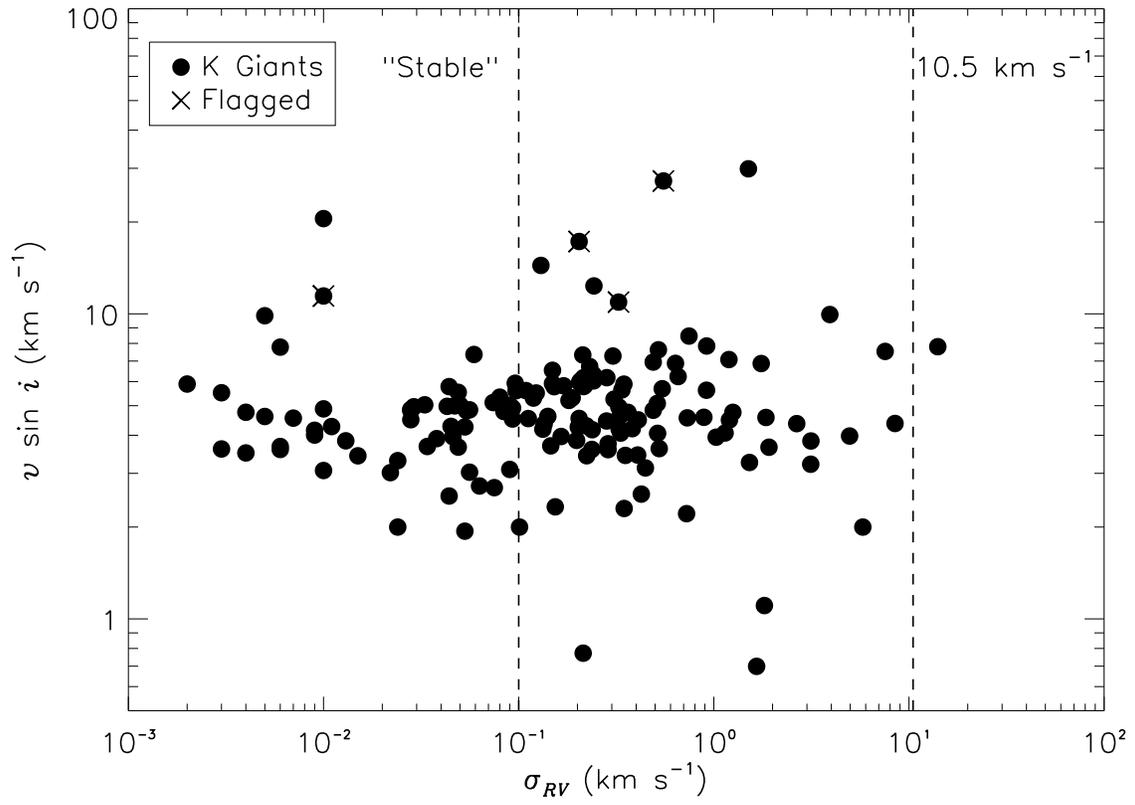}
\caption{ The projected rotational velocity compared to the radial velocity stability, $\sigma_{_{\rm RV}}$, for 151 stars in our sample that  have $\sigma_{_{\rm RV}}$  measured.  The radial velocity precision is 0.1~\kms.
\label{sigrv}}
\end{figure}

It is difficult to apply this result in a global sense because many variables (including the physical parameters of the system, the orientation of the system to an observer's line of sight, and the 
specific time of the observation) determine whether the velocity offset of the binary companions falls in the critical 10.5-15~\kms\ range.  
However, where we know the radial  velocity (RV) variability of a star, we can use it to
better quantify this effect.  A subset of our stars were observed for RV stability by B06; their survey could measure RV stability at the 0.1~\kms\ level.
In Figure \ref{sigrv}, we plot RV variations for the 151 stars in our sample with such data (eight of which are rapid rotators) against
their measured \vsini.  In this subsample,  56 (37\%) are considered stable ($\sigma_{_{\rm RV}} \leq 0.1$~\kms), and another 94 (62\%) are RV unstable but have variations less than 10.5~\kms\ (and thus are unlikely to be able to masquerade as rapid rotators).   The remaining  star, or  0.7\% \ of the sample, has an RV variability large enough to possibly create a broad blended feature; however,
recalling the discussion above,  this star has only a $9.5$\% chance that its companion is bright enough to have an effect.
Therefore, we expect that the total fraction of stars in our sample that appear to be rapid rotators but are in fact a binary with blended lines is only $0.06\%$,
which translates to approximately 1 of our rapid rotator stars.

\begin{figure} %F8
\includegraphics[scale=0.7]{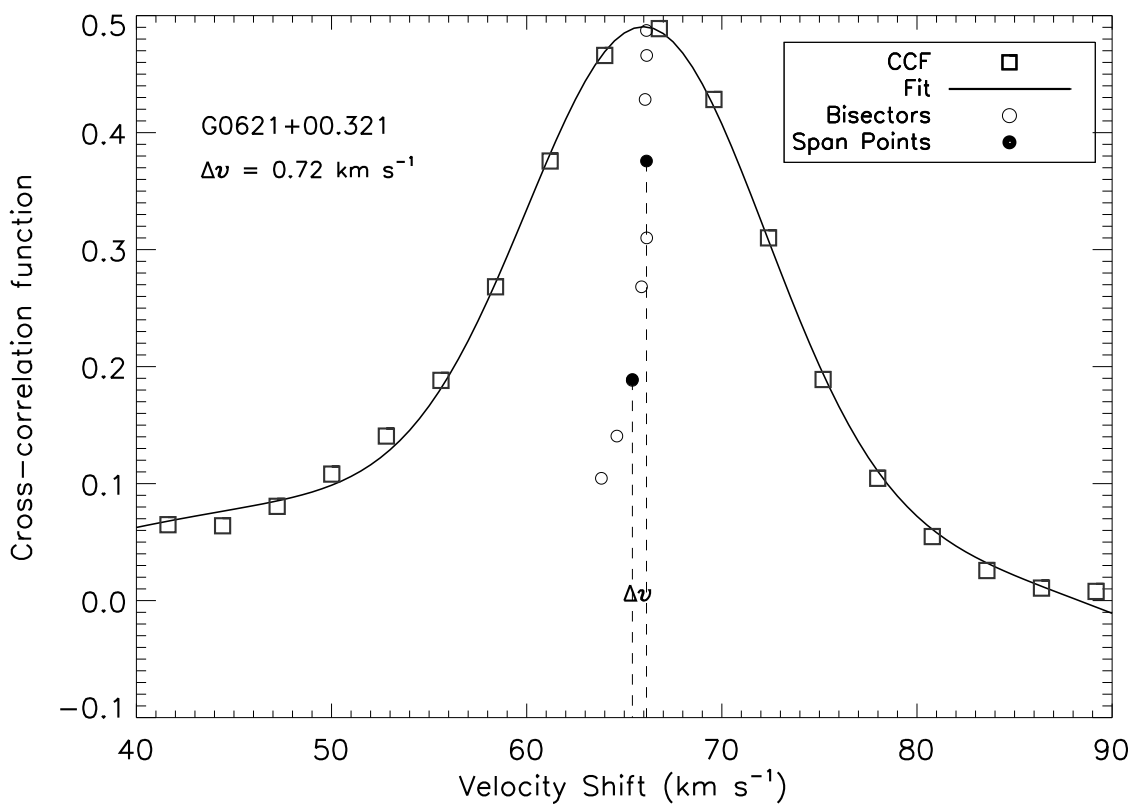}
\caption{Example of calculating the line-bisector velocity span of  the cross-correlation peak  for one echelle order of G0621+00.321.
The cross-correlation points are shown as squares, and a Gaussian fit to the cross-correlation peak is shown as the solid line. The bisector points are shown
as circles; the filled circles were used to compute the velocity span (see Section \ref{sec:binary} for details). \label{fig:ccfex}}
\end{figure}

Another approach to estimating the binary contamination is to try to directly detect the blended feature of a binary system with a line bisector method, since the blended lines would appear asymmetric for binary systems with stars of unequal temperature and/or brightness. Therefore, we attempted to measure line asymmetries
in the rapid rotator candidates.  Measuring asymmetries on many individual lines would be ideal, but at the S/N of our spectra,  only the strongest lines are visible above the noise. 
 Only  a handful  of our spectra have H$\alpha$, and although  all of the spectra have the sodium doublet, these lines suffer from line-of-sight interstellar medium absorption making them unsuitable for the analysis.
Instead, we tried  to assess line asymmetries in the cross-correlation peak itself by measuring the line-bisectors of the cross-correlation peaks on all of our rapid rotator candidates.  
 We computed a velocity span for each cross-correlation peak by finding the difference between the line center measured near  the top  of the line (87\% of the peak height) and mid-way down the line (55\%), as done in \cite{toner88}.   Figure \ref{fig:ccfex} shows an example of computing the line bisector points and velocity span.  We repeated this process on each echelle order for our spectra and found, unfortunately, that the order-to-order scatter for each star was too large to positively identify a line asymmetry with any statistical significance.
 
\section{Mechanisms for creating rapid rotators}
\label{sec:discuss}
As noted earlier, rapid rotation in giant stars is unusual, particularly for isolated stars that have no stellar companions with which
 to share angular momentum.   Although we have already found that there is a low probability of binary contamination where the lines of
 the two stars blend to mimic a rotationally broadened spectrum, binary companions can create legitimate rapid rotators.
 For an undetected companion star to bring about rapid rotation, it  must be close enough
to interact with the primary.  In particular, we address the situation where the star has been forced to co-rotate with its undetected companion,
and  we compute this for stars near both the RGB base and the RGB tip.
To have a \vsini \ of 10~\kms \ or larger, the period of co-rotation must be less than 
76 days ($\log P < 1.9$) for a star near the base of the RGB, assuming  a radius $\sim 15$~$R_{\sun}$.  For a star near the tip of the RGB, with $R \sim 100$~$R_{\sun}$, 
the period must be less than 500 days ($\log P < 2.7$) to be co-rotating at 10~\kms\ or more.  

Using the period distribution derived by \cite{duquennoy91}, the fraction of binary stars with periods less than 76 days and less than 500 days is approximately 10\% and 18\%, respectively.
From the study of \cite{famaey05}, the binary frequency among K giant stars is 13.7\%, and, accounting for randomized inclinations,
we find that the expected fraction of giant stars that would be rapidly rotating because of tidal interactions with a binary companion is between 1.3\% and 2.3\%.
 In our sample, we find that the rapid rotators comprise  2.2\% $\pm$ $0.4\%$, which implies that most  can be explained by tidally-locked binary stars.
The rapid rotators with the lowest surface gravities, and hence largest radii, are most likely to be tidally-locked binaries.  The tendency for the rapid rotators to be
more metal-poor may also have a similar explanation because at constant temperature, metal-poor giants are more luminous and have larger radii than more metal-rich giants.
 Curiously, the only known spectroscopic binary in our sample---Tyc3340-01195-1, introduced in Section \ref{sec:binary}---has \vsini$=9.5$~\kms, which is just shy of our rapid rotation definition.  However,
  its orbital period is at least twice the largest period for which tidally-locked rapid rotation is expected.  Its high \vsini\ still requires a non-binary star mechanism.

\begin{figure} %F9
\includegraphics[scale=0.7]{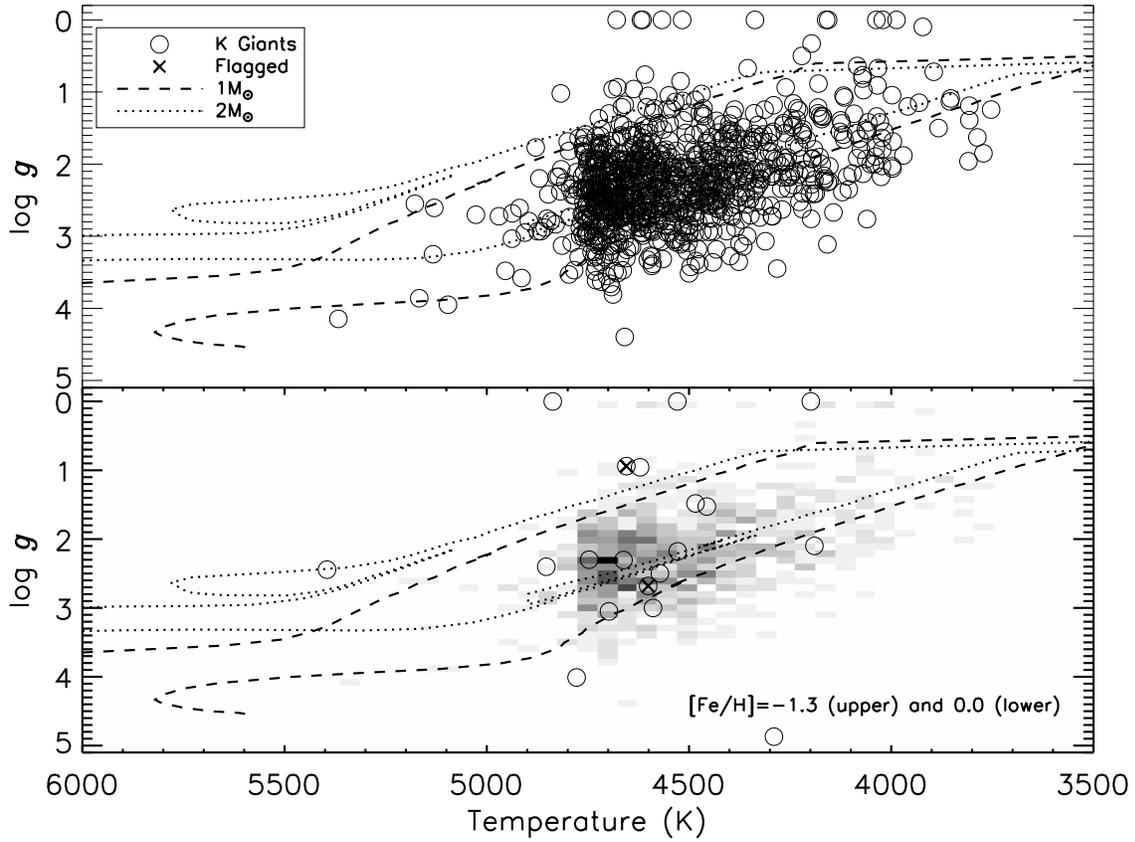}
\caption{Temperature-gravity plane for the normal stars (top) and rapid rotators (bottom).  Overplotted on each diagram
are two pairs of stellar evolution tracks \citep{giard00}: 1 \msun  (dotted) and 2 \msun (dashed) for [Fe/H]=-1.3 (upper pair) and [Fe/H]= 0.0 
(lower pair).  The number density distribution of the normal rotators in the top plot is plotted in grayscale in the bottom plot. \label{hr}}
\end{figure}

Our estimate of the number of expected rapid rotators from tidally-locked binaries does not exclude the possibility of rapid rotators without close binary companions.
  For these stars, one possible source of their excess angular momentum is from the stellar core \citep{simon89}. 
 If the core is differentially rotating and spinning much faster than the surface layers, then this high angular momentum 
reservoir may be tapped during first dredge-up.  Thus, we are interested in whether the normal and rapid rotators
populate different regions of the temperature-gravity plane (without distance information\footnote{We tried to obtain parallax measurements from the Hipparcos catalog to get distances, but because of the faintness of this sample, all of  target stars had parallax errors exceeding 11~mas. These large errors are generally comparable to or larger than the parallaxes themselves.}, we use surface gravity as a proxy for luminosity).  In Figure \ref{hr}, we plot  the temperature and $\log g$ of our K giants; 
the normal rotators are shown in the top panel and the rapid rotators in the lower panel.  Both plots include \cite{giard00} evolution tracks for solar and twice solar-mass stars
of both solar and sub-solar metallicities. The normal rotators 
are fairly evenly distributed over a wide range of temperatures and surface gravities, with a strongest concentration at a temperature
of about 4700 K and $\log g$ of 2.3 dex.  The rapid rotators are similarly distributed.  In particular, there does not seem to be any particular
clustering  in this diagram that would suggest that the rapid rotators are at the RGB bump. 
During the RGB bump evolutionary stage, the hydrogen-burning shell reaches the chemical discontinuity left behind by the retreated convection
zone; the star doubles back in its evolutionary track in a either the color-magnitude or
temperature-gravity plane, causing an  apparent over-density or bump. 
Extra mixing processes that are thought to occur at this phase, such as thermohaline mixing \citep{charbonnel07},  could possibly redistribute
angular momentum.   However, we conclude that this mechanism is not significantly affecting our sample
  because there is no clustering of the rapid rotators at the RGB bump.

Alternatively, external angular momentum from a sub-stellar companion could also cause rapid rotation in giant stars. Tidal forces can cause  a planet to spiral in towards the star,
depositing the orbital angular momentum into the stellar convection envelope.
\cite{carlberg09} found that such a signature is more likely to occur on the lower RGB.  However, the base of the RGB
generally occurs around 5100~K for solar metallicity stars (and  at higher temperatures for more metal-poor stars), and  it is clear from Figure \ref{hr} that 
this evolutionary stage, where finding rapid rotation from planet accretion is most probable, is not thoroughly probed by our K giant sample.
Another difficulty of the planet accretion explanation for rapid rotation is the fact that the rapid rotators seem to be more metal-poor on average than the normal rotators (Section \ref{sec:results}).  While there is an indication that giant stars with planets do not show the same tendency to be metal-rich as main sequence stars with planets do \citep{pasquini07,takeda08,ghezzi10},
 there is  as yet no reason to expect them to be more metal-poor.

In summary, our rapid rotators probably  gained their angular momentum in one of three ways.  First, they may be co-rotating with an unseen binary companion.  The probability of this scenario
is expected to increase as the star evolves because as the stellar radius increases, longer co-rotation periods (i.e., more distant companions) can cause \vsini\ to exceed 10 \kms.
Second, angular momentum redistribution within the giant star could have occurred.  However, the rarity  of rapid rotators implies that either all giants go through only a short-lived  phase of evolution where rapid rotation occurs or that only some giant stars experience unusually rapid rotation. In either case, we would expect an internal cause for the rapid rotation to lead to a clustering  of rapid rotators at a similar evolutionary stage, such as in Figure \ref{hr}, which we do not see.  Finally, angular momentum gained from a planetary companion is also considered. In contrast to the stellar companion scenario, planet-induced rapid rotation  is most likely to be seen on the lower red giant branch because the much smaller angular momentum reservoir of a planet  has the greatest chance of creating rapid rotation when the star's radius (and moment of inertia) is smallest.

\section{Conclusions}
\label{sec:conclusion}
We have developed a technique to estimate \vsini \ from high resolution, low S/N spectra using cross-correlation against a high quality template followed by a decomposition of the line broadening
components.  Applying this method to almost 1300 K giant stars drawn from the SIM Astrometric Grid Candidates, we have found 28 new candidate rapid rotators.
Of these 28, only  four have potentially unreliable \vsini \ estimates due to too low S/N in the spectra.   
One of the stars in our sample, Tyc5854-01526-1,
has an extremely high rotation at \vsini~$= 86.4$~\kms; such a high \vsini\ has only been seen in a handful of other red giant stars.

Because binary companions cannot be confidently identified and only one of our stars matched any binary star catalogs we queried, we evaluated two different ways that binary systems could be influencing our results.  We found that at most
one of our candidate rapid rotators could be a double lined spectroscopic binary where the lines of the component stars are blended so that they appear as a single set of
rotationally broadened lines.  Many of our rapid rotators are likely to be  binary star primaries that have been forced to co-rotate with their companions. 
An analysis of binary properties leads us to expect to find 24-32  tidally-locked binary-star rapid rotators in our sample, compared to the 28 that we did find.
The probability of a star having a companion close enough to tidally lock the star increases with larger stellar radius (and hence lower surface gravity for a given mass).
Unfortunately, we cannot presently determine which of our rapid rotators candidates are in fact in tidally-locked binaries. 

With the binary caveat in mind we have compared the rapid rotators to the normal rotator counterparts and  do not find any appreciable difference in the distribution of their temperatures. However, 
the rapid rotators do seem to have a slight over-abundance of low $\log g$ stars and tend to  be more metal-poor.   Low surface gravity stars in our rapid rotator sample are most likely to be
tidally-locked binaries because of their larger radii and increased likelihood of tidal interaction with a companion. Similarly, because metal-poor stars at a given
temperature are more luminous and have larger radii than more metal-rich stars, tidally-locked binaries could also explain the tendency of rapid rotators to be metal-poor.

We have also discussed two different formation scenarios for {\it isolated} rapid rotators.  The first is dredge up of angular momentum from a rapidly rotating stellar core. This mechanism is expected to cause rapid rotation at 
a particular phase in giant star evolution, but a clustering of the rapid rotators at this phase is not seen in our stars.
Other rapid rotators may have been formed  when  giants accreted angular momentum from a planetary companion; however, these are likely to be the most metal-rich
stars near the base of the RGB, which are stars not well represented in our sample.  Thus, we have no obvious explanation for the difference we see between the metallicity distributions of the rapid and slow rotators if they are not in binary systems.

Our survey search for rapid rotators is the faintest to date, and as such covers a more distant population of stars that may be older and more metal poor than stars in other surveys.
 Nevertheless, the total number of rapid rotators found constitutes 2.2\% of our sample and is consistent with other surveys of red giant stars. However, if our estimates of the number of rapid rotators that are simply co-rotating with a close stellar companion are accurate, then the number of {\it isolated} rapid rotators in our sample is much smaller than that of other surveys. This suggests that metal poor populations create fewer isolated rapid rotators.

\acknowledgments
This work has been supported by NASA/JPL through the Space Interferometry Mission Preparatory Science Grants 1201670 and 1222563, as well as NASA/JPL 
grant NRA-99-04-OSS-058.  JKC also acknowledges financial  support by NASA Headquarters under the NASA Earth and Space Science Fellowship Program for project 
08-Astro08F- 0012 and by the F. H. Levinson Fund of the Peninsula  Community Foundation. JKC would also like to thank Gail Zasowski for many helpful comments 
on earlier drafts of this manuscript and the anonymous referee for constructive comments that helped improve this paper.

\begin{deluxetable}{lrrrl}
\centering
\tablewidth{0pc}
\tabletypesize{\scriptsize}
\tablecaption{Program stars and aliases}
\tablecolumns{8}
\tablehead{
	\colhead{Name}&
	\colhead{Tycho}&
	\colhead{HIP}&
	\colhead{HD}&
	\colhead{BD}
}
\startdata
G2358+00.92         & Tyc0001-00976-1       &          &&              \\
Tyc0001-00818-1     & Tyc0001-00818-1       &     & 295&      BD-00    1 \\
Tyc0003-00199-1     & Tyc0003-00199-1      &     &  &    BD+01   64 \\
Tyc0003-00509-1     & Tyc0003-00509-1     &          & &          \\
Tyc0009-00104-1     & Tyc0009-00104-1       &         & & BD+05   58 \\
Tyc0016-00769-1     & Tyc0016-00769-1       &        & &  BD+04   74\\
Tyc0016-00837-1     & Tyc0016-00837-1       &        & &  BD+06   80\\
Tyc0018-00068-1     & Tyc0018-00068-1       &       & &  BD+06  132 \\
Tyc0018-00533-1     & Tyc0018-00533-1    &4325 & &   BD+04  138\\
\enddata
\tablecomments{ This table is available in its entirety as a plain text ancillary file associated with the arXiv article. \label{tab:alias}}
\end{deluxetable}

\begin{deluxetable}{lrrrccc}
\centering
\tablewidth{0pc}
\tabletypesize{\scriptsize}
\tablecaption{Properties of stars not flagged in the \vsini\ derivation. \label{tab1}}
\tablecolumns{8}
\tablehead{
	\colhead{Object Name}&
	\colhead{R.A.}&
	\colhead{Dec.}&
	\colhead{\vsini}&
	\colhead{Error}&
	\colhead{N pts.}&
	\colhead{Temperature}\\
	\colhead{}&
	\colhead{(J2000.0)}&
	\colhead{(J2000.0)}&
	\colhead{(\kms)}&
	\colhead{(\kms)}&
	\colhead{}&
	\colhead{(K)}
}
\startdata
G0000+67.16328 & 23:58:53.52 & $+67$:34:02.04 & 5.2 & 0.4 & 22 & 4545 \\ 
G0001+00.94 & 00:04:20.08 &$ +00$:27:12.25 & 3.2 & 0.3 & 24 & 4446 \\ 
G0011+05.87 & 00:14:18.85 & $+05$:57:37.52 & 8.2 & 0.3 & 22 & 4643 \\ 
G0011+16.129 & 00:14:58.70 & $+17$:05:04.22 & 3.6 & 0.3 & 22 & 4482 \\ 
G0011+16.75 & 00:14:26.65 & $+17$:16:03.00 & 2.6 & 0.3 & 21 & 4813 \\ 
G0021+00.24 & 00:23:09.34 & $+00$:07:42.77 & 40.4 & 0.0 & 21 & 4976 \\ 
G0022+00.21 & 00:24:30.62 & $+00$:18:13.37 &3.7 & 0.0 & 21 & 5026 \\ 
G0024+61.2521 & 00:24:10.50 & $+62$:26:51.90& 5.2 & 0.3 & 21 & 4346 \\ 
G0024+61.6685 & 00:24:35.49 & $+62$:18:14.85 & 4.1 & 0.3 & 22 & 4508 \\ 
G0024+61.9144 & 00:29:17.09 & $+62$:13:14.82 & 6.0 & 0.4 & 22 & 4592 \\ 
G0032-05.63 & 00:35:13.06 & $-05$:29:09.04 & $<3.0$ & 0.7 &\ 7 & 5062 \\ 
G0043+00.77 & 00:46:30.97 & $+00$:18:30.29 & $<3.0$  & 0.4 & 19 & 5060 \\ 
G0046+00.42 & 00:48:57.36 &$ +00$:21:58.86 & 1.3 & 0.4 & 21 & 4712 \\ 
G0054+05.74 & 00:57:49.44 & $+05$:55:54.82& 3.0 & 0.5 & 20 & 4998 \\ 
G0108+00.67 & 01:11:42.07 & $+00$:12:50.60 & 6.2 & 1.1 &\ 5 & 4429 \\ 
\enddata
\tablecomments{ This table is available in its entirety as a plain text ancillary file associated with the arXiv article. }
\end{deluxetable}

\begin{deluxetable}{lrrrccc}
\centering
\tablewidth{0pc}
\tabletypesize{\scriptsize}
\tablecaption{Properties of stars that were flagged in the \vsini\ derivation. \label{tabFL}}
\tablecolumns{8}
\tablehead{
	\colhead{Object Name}&
	\colhead{R.A.}&
	\colhead{Dec.}&
	\colhead{\vsini}&
	\colhead{Error}&
	\colhead{N pts.}&
	\colhead{Temperature}\\
	\colhead{}&
	\colhead{(J2000.0)}&
	\colhead{(J2000.0)}&
	\colhead{(\kms)}&
	\colhead{(\kms)}&
	\colhead{}&
	\colhead{(K)}
}
\startdata
G0335+05.146 & 03:38:17.56 & $+05$:51:22.54& 4.5 & 0.5 & 34 & 5145 \\ 
G0505+05.506 & 05:08:19.36 & $+05:$38:18.25& 17.1 & 3.0 & 37 & 4815 \\ 
G0632+05.983 & 06:35:21.77 & $+05$:30:52.44& 5.1 & 0.9 & 16 & 4751 \\ 
G0633-05.181 & 06:35:28.19 & $-05$:45:08.37& 12.8 & 2.0 & 17 & 3917 \\ 
G0900+00.55 & 09:02:08.54 & $-00$:22:27.14& 4.1 & 0.6 & 16 & 4623 \\ 
G0907-11.52 & 09:09:06.66 & $-11$:37:15.97& 6.7 & 1.6 & 12 & 4554 \\ 
G1333-16.35 & 13:35:08.68 & $-17$:02:06.54 & 4.8 & 1.0 & 17 & 4760 \\ 
G1342+05.77 & 13:45:50.45 & $+05$:13:06.63 & 10.8 & 0.8 & 40 & 4747 \\ 
G1800+00.410 & 18:02:13.45 & $+00$:03:02.37& 5.4 & 0.6 & 18 & 4378 \\ 
G1819-16.313 & 18:21:23.87 & $-16$:39:23.47 & 12.2 & 1.5 & 70 & 4367 \\ 
G1838+28.14307 & 18:40:34.30 & $+27$:56:17.99 & 4.7 & 0.8 & 20 & 4032 \\ 
G2237-16.2017 & 22:39:41.87 & $-16$:35:45.70 & 6.4 & 0.8 & 15 & 4368 \\ 
G2326-05.7160 & 23:29:21.37 & $-05$:26:32.55& 4.3 & 0.7 & 21 & 4912 \\ 
\enddata
\end{deluxetable}

\end{document}